\input epsf
\newfam\scrfam
\batchmode\font\tenscr=rsfs10 \errorstopmode
\ifx\tenscr\nullfont
        \message{rsfs script font not available. Replacing with calligraphic.}
        \def\scr{\cal}
\else   
        \font\sevenscr=rsfs7
        \font\fivescr=rsfs5
        \skewchar\tenscr='177 \skewchar\sevenscr='177 \skewchar\fivescr='177
        \textfont\scrfam=\tenscr \scriptfont\scrfam=\sevenscr
        \scriptscriptfont\scrfam=\fivescr
        \def\scr{\fam\scrfam}
        \def\cal{\scr}
\fi
\catcode`\@=11
\newfam\frakfam
\batchmode\font\tenfrak=eufm10 \errorstopmode
\ifx\tenfrak\nullfont
        \message{eufm font not available. Replacing with italic.}
        
\else
	
	\font\sevenfrak=eufm7 \font\fivefrak=eufm5
	\textfont\frakfam=\tenfrak
	\scriptfont\frakfam=\sevenfrak \scriptscriptfont\frakfam=\fivefrak
	
\fi
\catcode`\@=\active
\newfam\msbfam
\batchmode\font\twelvemsb=msbm10 scaled\magstep1 \errorstopmode
\ifx\twelvemsb\nullfont\def\Bbb{\bf}

	\message{Blackboard bold not available. Replacing with boldface.}
\else   \catcode`\@=11
        \font\tenmsb=msbm10 \font\sevenmsb=msbm7 \font\fivemsb=msbm5
        \textfont\msbfam=\tenmsb
        \scriptfont\msbfam=\sevenmsb \scriptscriptfont\msbfam=\fivemsb
        \def\Bbb{\relax\expandafter\Bbb@}
        \def\Bbb@#1{{\Bbb@@{#1}}}
        \def\Bbb@@#1{\fam\msbfam\relax#1}
        \catcode`\@=\active

\fi
        \font\eightrm=cmr8              \def\xrm{\eightrm}
        \font\eightbf=cmbx8             \def\xbf{\eightbf}
        \font\eightit=cmti10 at 8pt     \def\xit{\eightit}
                     
        \font\eightcp=cmcsc8
        \font\eighti=cmmi8              \def\xold{\eighti}
        \font\eightib=cmmib8             \def\xbold{\eightib}
        \font\teni=cmmi10               \def\old{\teni}
        \font\tencp=cmcsc10

        \font\twelvecp=cmcsc10 scaled\magstep1

	 at10pt	
	\font\twelvehelvbold=phvb at12pt
	 at14pt
	\font\sixteenhelvbold=phvb at16pt

\def\noblackbox{\overfullrule=0pt}
\noblackbox

\newtoks\headtext
\headline={\ifnum\pageno=1\hfill\else
	\ifodd\pageno{\eightcp\the\headtext}{ }\dotfill{ }{\old\folio}
	\else{\old\folio}{ }\dotfill{ }{\eightcp\the\headtext}\fi
	\fi}
\def\makeheadline{\vbox to 0pt{\vss\noindent\the\headline\break
\hbox to\hsize{\hfill}}
        \vskip2\baselineskip}
\newcount\infootnote
\infootnote=0
\def\foot#1#2{\infootnote=1
\footnote{${}^{#1}$}{\vtop{\baselineskip=.75\baselineskip
\advance\hsize by -\parindent\noindent{\xrm #2}}}\infootnote=0$\,$}
\newcount\refcount
\refcount=1
\newwrite\refwrite
\def\oldsize{\ifnum\infootnote=1\xold\else\old\fi}
\def\ref#1#2{
	\def#1{{{\oldsize\the\refcount}}\ifnum\the\refcount=1\immediate\openout\refwrite=\jobname.refs\fi\immediate\write\refwrite{\item{[{\xold\the\refcount}]} 
	#2\hfill\par\vskip-2pt}\xdef#1{{\noexpand\oldsize\the\refcount}}\global\advance\refcount by 1}
	}
\def\refout{\catcode`\@=11
        \xrm\immediate\closeout\refwrite
        \vskip2\baselineskip
        {\noindent\twelvecp References}\hfill\vskip\baselineskip
        \baselineskip=.75\baselineskip
        \input\jobname.refs
        \baselineskip=4\baselineskip \divide\baselineskip by 3
        \catcode`\@=\active\rm}

\def\hepth#1{\href{http://xxx.lanl.gov/abs/hep-th/#1}{hep-th/{\xold#1}}}

\def\arxiv#1#2{\href{http://arxiv.org/abs/#1.#2}{arXiv:{\xold#1}.{\xold#2}}}
\def\jhep#1#2#3#4{\href{http://jhep.sissa.it/stdsearch?paper=#2\%28#3\%29#4}{J. High Energy Phys. {\xbold #1#2} ({\xold#3}) {\xold#4}}}

\def\ATMP#1#2#3{Adv. Theor. Math. Phys. {\xbold#1} ({\xold#2}) {\xold#3}}

\def\CQG#1#2#3{Class. Quantum Grav. {\xbold#1} ({\xold#2}) {\xold#3}}

\def\JHEP{\jhep}

\def\LMP#1#2#3{Lett. Math. Phys. {\xbold#1} ({\xold#2}) {\xold#3}}

\def\NPB#1#2#3{Nucl. Phys. {\xbf B}{\xbold#1} ({\xold#2}) {\xold#3}}

\def\PLB#1#2#3{Phys. Lett. {\xbf B}{\xbold#1} ({\xold#2}) {\xold#3}}

\def\PRD#1#2#3{Phys. Rev. {\xbf D}{\xbold#1} ({\xold#2}) {\xold#3}}

\newcount\sectioncount
\sectioncount=0
\def\section#1#2{\global\eqcount=0
	\global\subsectioncount=0
        \global\advance\sectioncount by 1
	\ifnum\sectioncount>1
	        \vskip2\baselineskip
	\fi
\line{\twelvecp\the\sectioncount. #2\hfill}
       \vskip.5\baselineskip\noindent
        \xdef#1{{\old\the\sectioncount}}}
\newcount\subsectioncount
\def\subsection#1#2{\global\advance\subsectioncount by 1
	\vskip.75\baselineskip\noindent
\line{\tencp\the\sectioncount.\the\subsectioncount. #2\hfill}
	\vskip.5\baselineskip\noindent
	\xdef#1{{\old\the\sectioncount}.{\old\the\subsectioncount}}}
\def\immediatesubsection#1#2{\global\advance\subsectioncount by 1
\vskip-\baselineskip\noindent
\line{\tencp\the\sectioncount.\the\subsectioncount. #2\hfill}
	\vskip.5\baselineskip\noindent
	\xdef#1{{\old\the\sectioncount}.{\old\the\subsectioncount}}}
\newcount\appendixcount
\appendixcount=0
\def\appendix#1{\global\eqcount=0
        \global\advance\appendixcount by 1
        \vskip2\baselineskip\noindent
        \ifnum\the\appendixcount=1
        \hbox{\twelvecp Appendix A: #1\hfill}\vskip\baselineskip\noindent\fi
    \ifnum\the\appendixcount=2
        \hbox{\twelvecp Appendix B: #1\hfill}\vskip\baselineskip\noindent\fi
    \ifnum\the\appendixcount=3
        \hbox{\twelvecp Appendix C: #1\hfill}\vskip\baselineskip\noindent\fi}
\def\acknowledgements{\vskip2\baselineskip\noindent
        \underbar{\it Acknowledgements:}\ }
\newcount\eqcount
\eqcount=0
\def\Eqn#1{\global\advance\eqcount by 1
\ifnum\the\sectioncount=0
	\xdef#1{{\old\the\eqcount}}
	\eqno({\oldstyle\the\eqcount})
\else
        \ifnum\the\appendixcount=0
	        \xdef#1{{\old\the\sectioncount}.{\old\the\eqcount}}
                \eqno({\oldstyle\the\sectioncount}.{\oldstyle\the\eqcount})\fi
        \ifnum\the\appendixcount=1
	        \xdef#1{{\oldstyle A}.{\old\the\eqcount}}
                \eqno({\oldstyle A}.{\oldstyle\the\eqcount})\fi
        \ifnum\the\appendixcount=2
	        \xdef#1{{\oldstyle B}.{\old\the\eqcount}}
                \eqno({\oldstyle B}.{\oldstyle\the\eqcount})\fi
        \ifnum\the\appendixcount=3
	        \xdef#1{{\oldstyle C}.{\old\the\eqcount}}
                \eqno({\oldstyle C}.{\oldstyle\the\eqcount})\fi
\fi}
\def\eqn{\global\advance\eqcount by 1
\ifnum\the\sectioncount=0
	\eqno({\oldstyle\the\eqcount})
\else
        \ifnum\the\appendixcount=0
                \eqno({\oldstyle\the\sectioncount}.{\oldstyle\the\eqcount})\fi
        \ifnum\the\appendixcount=1
                \eqno({\oldstyle A}.{\oldstyle\the\eqcount})\fi
        \ifnum\the\appendixcount=2
                \eqno({\oldstyle B}.{\oldstyle\the\eqcount})\fi
        \ifnum\the\appendixcount=3
                \eqno({\oldstyle C}.{\oldstyle\the\eqcount})\fi
\fi}
\def\multi{\global\advance\eqcount by 1}
\def\multieq#1#2{\xdef#1{{\old\the\eqcount#2}}
        \eqno{({\oldstyle\the\eqcount#2})}}
\newtoks\url
\def\Href#1#2{\catcode`\#=12\url={#1}\catcode`\#=\active#2}
\def\href#1#2{{#2}}

\parskip=3.5pt plus .3pt minus .3pt
\baselineskip=14pt plus .1pt minus .05pt
\lineskip=.5pt plus .05pt minus .05pt
\lineskiplimit=.5pt
\abovedisplayskip=18pt plus 4pt minus 2pt
\belowdisplayskip=\abovedisplayskip
\hsize=14cm
\vsize=19cm
\hoffset=1.5cm
\voffset=1.8cm
\frenchspacing
\footline={}
\raggedbottom
\def\ts{\textstyle}
\def\ss{\scriptstyle}
\def\sss{\scriptscriptstyle}
\def\*{\partial}
\def\punkt{\,\,.}
\def\komma{\,\,,}

\def\={\!=\!}
\def\small#1{{\hbox{$#1$}}}

\def\fraction#1{\small{1\over#1}}
\def\fr{\fraction}
\def\Fraction#1#2{\small{#1\over#2}}
\def\Fr{\Fraction}

\def\eg{{\tenit e.g.}}

\def\ie{{\tenit i.e.}}

\def\a{\alpha}
\def\b{\beta}

\def\d{\delta}
\def\e{\varepsilon}
\def\g{\gamma}
\def\l{\lambda}

\def\s{\sigma}
\def\th{\theta}

\def\ra{\rightarrow}

\def\II{I\hskip-.8pt I}

\def\ra{\rightarrow}
\def\la{\leftarrow}

\def\rarrowover#1{\vtop{\baselineskip=0pt\lineskip=0pt
      \ialign{\hfill##\hfill\cr$\ra$\cr$#1$\cr}}}

\def\larrowover#1{\vtop{\baselineskip=0pt\lineskip=0pt
      \ialign{\hfill##\hfill\cr$\la$\cr$#1$\cr}}}


\def\modprod#1{\raise0pt\vtop{\baselineskip=0pt\lineskip=0pt
      \ialign{\hfill##\hfill\cr$\circ$\cr${\sss #1}$\cr}}}

\def\axi{\xi^{\star}}


\ref\BaggerLambertI{J. Bagger and N. Lambert, {\xit ``Modeling
multiple M2's''}, \PRD{75}{2007}{045020} [\hepth{0611108}].}

\ref\BaggerLambertII{J. Bagger and N. Lambert, {\xit ``Gauge symmetry
and supersymmetry of multiple M2-branes''}, \PRD{77}{2008}{065008}
[\arxiv{0711}{0955}].} 

\ref\Gustavsson{A. Gustavsson, {\xit ``Algebraic structures on
parallel M2-branes''}, \arxiv{0709}{1260}.}

\ref\CederwallNilssonTsimpisI{M. Cederwall, B.E.W. Nilsson and D. Tsimpis,
{\xit ``The structure of maximally supersymmetric super-Yang--Mills
theory---constraining higher order corrections''},
\jhep{01}{06}{2001}{034} 
[\hepth{0102009}].}

\ref\CederwallNilssonTsimpisII{M. Cederwall, B.E.W. Nilsson and D. Tsimpis,
{\xit ``D=10 super-Yang--Mills at $\ss O(\a'^2)$''},
\JHEP{01}{07}{2001}{042} [\hepth{0104236}].}

\ref\BerkovitsParticle{N. Berkovits, {\xit ``Covariant quantization of
the superparticle using pure spinors''}, \jhep{01}{09}{2001}{016}
[\hepth{0105050}].}

\ref\SpinorialCohomology{M. Cederwall, B.E.W. Nilsson and D. Tsimpis,
{\xit ``Spinorial cohomology and maximally supersymmetric theories''},
\jhep{02}{02}{2002}{009} [\hepth{0110069}];
M. Cederwall, {\xit ``Superspace methods in string theory, supergravity and gauge theory''}, Lectures at the XXXVII Winter School in Theoretical Physics ``New Developments in Fundamental Interactions Theories'',  Karpacz, Poland,  Feb. 6-15, 2001, \hepth{0105176}.}

\ref\Movshev{M. Movshev and A. Schwarz, {\xit ``On maximally
supersymmetric Yang--Mills theories''}, \NPB{681}{2004}{324}
[\hepth{0311132}].}

\ref\BerkovitsI{N. Berkovits,
{\xit ``Super-Poincar\'e covariant quantization of the superstring''},
\jhep{00}{04}{2000}{018} [\hepth{0001035}].}

\ref\BerkovitsNonMinimal{N. Berkovits,
{\xit ``Pure spinor formalism as an N=2 topological string''},
\jhep{05}{10}{2005}{089} [\hepth{0509120}].}

\ref\CederwallNilssonSix{M. Cederwall and B.E.W. Nilsson, {\xit ``Pure
spinors and D=6 super-Yang--Mills''}, \arxiv{0801}{1428}.}

\ref\CGNN{M. Cederwall, U. Gran, M. Nielsen and B.E.W. Nilsson,
{\xit ``Manifestly supersymmetric M-theory''},
\JHEP{00}{10}{2000}{041} [\hepth{0007035}];
{\xit ``Generalised 11-dimensional supergravity''}, \hepth{0010042}.
}

\ref\CGNT{M. Cederwall, U. Gran, B.E.W. Nilsson and D. Tsimpis,
{\xit ``Supersymmetric corrections to eleven-dimen\-sional supergravity''},
\jhep{05}{05}{2005}{052} [\hepth{0409107}].}

\ref\NilssonPure{B.E.W.~Nilsson,
{\xit ``Pure spinors as auxiliary fields in the ten-dimensional
supersymmetric Yang--Mills theory''},
\CQG3{1986}{{\xrm L}41}.}

\ref\HowePureI{P.S. Howe, {\xit ``Pure spinor lines in superspace and
ten-dimensional supersymmetric theories''}, \PLB{258}{1991}{141}.}

\ref\HowePureII{P.S. Howe, {\xit ``Pure spinors, function superspaces
and supergravity theories in ten and eleven dimensions''},
\PLB{273}{1991}{90}.} 

\ref\CederwallBLG{M. Cederwall, {\xit ``N=8 superfield formulation of
the Bagger--Lambert--Gustavsson model''}, \jhep{08}{09}{2008}{116}
[\arxiv{0808}{3242}].}

\ref\CederwallABJM{M. Cederwall, {\xit ``Superfield actions for N=8 
and N=6 conformal theories in three dimensions''},
\jhep{08}{10}{2008}{70}
[\arxiv{0808}{3242}].}

\ref\MarneliusOgren{R. Marnelius and M. \"Ogren, {\xit ``Symmetric
inner products for physical states in BRST quantization''},
\NPB{351}{1991}{474}.} 

\ref\BerkovitsICTP{N. Berkovits, {\xit ``ICTP lectures on covariant
quantization of the superstring''}, proceedings of the ICTP Spring
School on Superstrings and Related Matters, Trieste, Italy, 2002
[\hepth{0209059}.]} 

\ref\ABJM{O. Aharony, O. Bergman, D.L. Jafferis and J. Maldacena,
{\xit ``N=6 superconformal Chern--Simons-matter theories, M2-branes
and their gravity duals''}, \arxiv{0806}{1218}.}

\ref\ElevenSG{E. Cremmer, B. Julia and J. Sherk, 
{\xit ``Supergravity theory in eleven-dimensions''},
\PLB{76}{1978}{409}.}

\ref\ElevenSGSuperspace{L. Brink and P. Howe, 
{\xit ``Eleven-dimensional supergravity on the mass-shell in superspace''},
\PLB{91}{1980}{384};
E. Cremmer and S. Ferrara,
{\xit ``Formulation of eleven-dimensional supergravity in superspace''},
\PLB{91}{1980}{61}.}
 
\ref\BatalinVilkovisky{I.A. Batalin and G.I. Vilkovisky, {\xit ``Gauge
algebra and quantization''}, \PLB{102}{1981}{27}.}

\ref\FusterBVReview{A. Fuster, M. Henneaux and A. Maas, {\xit
``BRST-antifield quantization: a short review''}, \hepth{0506098}.}

\ref\CederwallInProgress{M. Cederwall, work in progress.}

\ref\ZwiebachClosedBV{B. Zwiebach, {\xit ``Closed string field theory:
    Quantum action and the BV master equation''}, \hepth{9206084}.}

\ref\BerkovitsMembrane{N. Berkovits,
	{\xit ``Towards covariant quantization of the supermembrane''},
	\JHEP{02}{09}{2002}{051} [\hepth{0201151}].}

\ref\BerkovitsNekrasovCharacter{N. Berkovits and N. Nekrasov, {\xit
    ``The character of pure spinors''}, \LMP{74}{2005}{75}
  [\hepth{0503075}].}

\ref\BerkovitsNekrasovMultiloop{N. Berkovits and N. Nekrasov, {\xit
    ``Multiloop superstring amplitudes from non-minimal pure spinor
    formalism''}, \jhep{06}{12}{2006}{029} [\hepth{0609012}].}

\ref\HoweWeyl{P. Howe, {\xit ``Weyl superspace''},
  \PLB{415}{1997}{149} [\hepth{9707184}].}

\ref\BoulangerUniqueness{N. Boulanger, T. Damour, L. Gualtieri and
  M. Henneaux, {\xit ``Inconsistency of interacting, multigraviton
    theories''}, \NPB{597}{2001}{127} [\hepth{0007220}].}

\ref\GAMMA{U. Gran,
{\xit ``GAMMA: A Mathematica package for performing gamma-matrix 
algebra and Fierz transformations in arbitrary dimensions''},
\hepth{0105086}.}

\ref\AnguelovaGrassiVanhove{L. Anguelova, P.A. Grassi and P. Vanhove,
  {\xit ``Covariant one-loop amplitudes in D=11''},
  \NPB{702}{2004}{269} [\hepth{0408171}].}

\ref\GrassiVanhove{P.A. Grassi and P. Vanhove, {\xit ``Topological M
    theory from pure spinor formalism''}, \ATMP{9}{2005}{285}
  [\hepth{0411167}].}


\headtext={M. Cederwall: ``Towards a manifestly supersymmetric...''}

\line{
\epsfxsize=18mm
\epsffile{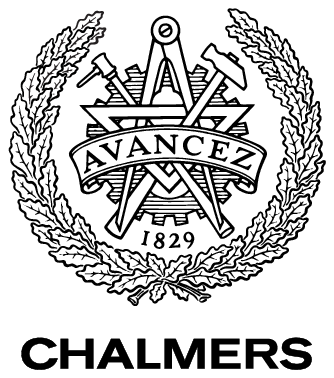}
\hfill}
\vskip-12mm
\line{\hfill G\"oteborg preprint}
\line{\hfill December, {\old2009}}
\line{\hrulefill}

\vfill
\vskip.5cm

\centerline{\sixteenhelvbold
Towards a manifestly supersymmetric action } 

\vskip4\parskip

\centerline{\sixteenhelvbold
for 11-dimensional supergravity} 

\vfill

\centerline{\twelvehelvbold
Martin Cederwall}

\vfill

\centerline{\it Fundamental Physics}
\centerline{\it Chalmers University of Technology}
\centerline{\it SE 412 96 G\"oteborg, Sweden}

\vfill

{\narrower\noindent \underbar{Abstract:} We investigate the
possibility of writing a manifestly supersymmetric action for
11-dimensional supergravity. The construction
involves an explicit relation between the fields in the
super-vielbein and the super-3-form, and uses non-minimal pure
spinors. 
A simple cubic interaction term for a single scalar superfield is
found.
\smallskip}
\vfill

\font\xxtt=cmtt6

\vtop{\baselineskip=.6\baselineskip\xxtt
\line{\hrulefill}
\catcode`\@=11
\line{email: martin.cederwall@chalmers.se\hfill}
\catcode`\@=\active
}

\eject

\def\l{\lambda}
\def\lb{\bar\lambda}

\section\Intro{Introduction}Eleven-dimensional supergravity [\ElevenSG] is
important being the low-energy limit of M-theory, and hence of a
strong-coupling limit of string theory. It is the highest-dimensional
supersymmetric model including gravity, and gives rise to most
lower-dimensional supergravities. With 32 supercharges, it has
maximal supersymmetry and a traditional superspace description
[\ElevenSGSuperspace] puts the theory on-shell.

It has been known for some time that
pure spinor superfields provide a powerful
tool for formulating supersymmetric field and string theories
[\NilssonPure,\HowePureI,\HowePureII,\BerkovitsI,\BerkovitsParticle,\CGNN,\CederwallNilssonTsimpisI,\CederwallNilssonTsimpisII,\SpinorialCohomology,\Movshev,\CGNT,\BerkovitsNonMinimal,\CederwallNilssonSix,\CederwallBLG,\CederwallABJM]. 
This is
especially true in models with maximal supersymmetry, where the
on-shell closure of the supersymmetry acting on an ordinary superfield
is turned into an advantage---the constraint on the ordinary
superfield, which enforces the equations of motion, is encoded in a
cohomological equation of the type $Q\Psi+\ldots=0$, which is the equation of
motion for the pure spinor superfield.
Supermultiplets arise as cohomologies in pure
spinor superfields. It is striking that these cohomologies rely
only on the purely algebraic (bosonic) constraint structure of the
pure spinor. Not only do the physical fields arise in this way, but
also the full set of ghosts and antifields. Pure spinor superfield
theory inevitably leads to a Batalin--Vilkovisky (BV) formalism 
[\BatalinVilkovisky,\FusterBVReview].

Much of the work on pure spinors in connection with supersymmetry has
been done for strings, and less for supersymmetric field theories. The
main difference between the treatments of string theory and field
theory is that strings in principle are treated in a first-quantised
manner, with interactions represented by vertex operators and the
geometry of the world sheet, while theories with fundamental particle
excitations are full-fledged field theories. A few maximally
supersymmetric field theories have been formulated this way, including
$D=10$ super-Yang--Mills (and its dimensional reductions) 
[\BerkovitsParticle,\CederwallNilssonTsimpisI] and
the Bagger--Lambert--Gustavsson
[\BaggerLambertI,\Gustavsson,\BaggerLambertII] 
and
Aharony--Bergman--Jafferis--Maldacena [\ABJM] conformal models in $D=3$
[\CederwallBLG,\CederwallABJM],
although none of them have been used for systematic quantum calculations.

So far, no supersymmetric field theory containing gravity has been
formulated with manifest supersymmetry 
beyond the free level. One purpose of the present paper is
to examine such formulations. An obvious drawback will be that
manifest background invariance is sacrificed, since the form of the
BRST operator $Q$ will encode geometric data of the background around
which one chooses to expand. On the other hand, all supersymmetry is
maintained; our choice is to give this highest priority.

When deciding which supergravity model to try to give a
pure spinor superfield formulation, one would of course like the
simplest one possible. But on the other hand, maximal supersymmetry is
essential for simplicity. For a model with half-maximal or less
supersymmetry, the cohomology will give an off-shell multiplet, and
one will need yet more constraints on the fields to obtain the
equations of motion. This happens for example in $D=6$, $N=1$
super-Yang--Mills theory, where a second BRST operator   
effectively sets the auxiliary fields to zero
[\CederwallNilssonSix]. Similarly, in $D=10$,
$N=1$ supergravity, the cohomology gives a (partially) off-shell
supermultiplet [\CederwallInProgress]. 
Therefore we want to begin with a maximally
supersymmetric model, and maybe address the question of lower
supersymmetry once the maximal case is understood. 
The only candidates are type \II B supergravity
and $D=11$ supergravity and their dimensional reductions. Type \II B
has a self-dual field strength, which complicates the formulation of
an action (this is actually reflected in the cohomology, which due to
the absence of certain anti-fields does not yield a natural measure
[\CederwallInProgress]).
There is no toy model. $D=11$ supergravity seems to be essentially the
only choice.

Pure spinor superfield formulations tend to have some remarkable
properties, as an extra bonus in addition to the manifest
supersymmetry. 
The action for $D=10$ super-Yang--Mills is
Chern--Simons-like, and has only a cubic interaction
[\BerkovitsParticle]. 
The conformal
models in $D=3$, whose component actions contain couplings of six
scalar fields, simplify enormously in the pure spinor framework, where
the matter superfields only have a minimal coupling to the
Chern--Simons field [\CederwallBLG,\CederwallABJM]. 
Higher order interactions arise when auxiliary
fields are eliminated (in both cases the fermionic component of the gauge
connection on superspace). One may imagine that these simplifications
may turn out to be useful in quantum calculations, where Feynman
diagrams will be built with 3-vertices only.

It will be interesting to see to what extent something similar
happens for supergravity. There is of course no reason to believe that
the action will be polynomial, but there may be simplifications in the
series of interactions that makes the theory more tractable. In this
sense one may think of the supergravity action as a toy model for
closed string field theory [\ZwiebachClosedBV]. We will
not have much to say about this, but plan to investigate the issue in
the future.

The linearised cohomology giving $D=11$ supergravity is known 
[\CGNN,\SpinorialCohomology,\BerkovitsMembrane]. There
is a fermionic 
scalar field $\Psi$ of dimension $-3$ and ghost number $3$ whose lowest
component is the third order ghost for the tensor field. The physical
fields of ghost number 0 sit in the field as $\l^\a\l^\b\l^\g
C_{\a\b\g}(x,\th)$, where $\l^\a$ is the pure spinor and $C_{\a\b\g}$
the lowest-dimensional part of the superspace 3-form $C$. As will be
reviewed later, there is a natural measure on the pure spinor space,
and it is straightforward to write an action $\int\Psi Q\Psi$ 
giving the linearised 
equations of motion [\BerkovitsMembrane]. 
The integrand has ghost number 7 and dimension
$-6$. Clearly, since three powers of $\Psi$ already gives ghost number 9, 
some operators with negative ghost number have to be
introduced in the interaction terms. Partial results concerning
interactions have been obtained in
refs. [\BerkovitsMembrane,\AnguelovaGrassiVanhove,\GrassiVanhove].  

In the superfield treatment of $D=11$ supergravity
[\ElevenSGSuperspace], the linearised
fields can be obtained not only from the 3-form, but also from the
super-vielbein. This is reflected in the existence of another pure
spinor field $\Phi^a$, where $a$ is a vector index. This field is
fermionic, has ghost
number 1 and
dimension $-1$ and starts with
the diffeomorphism ghost. The physical fields sit in $\Phi^a$ as $\l^\a
E_\a{}^a(x,\th)$, where $E_\a{}^a$ is (part of) the linearised
lowest-dimensional component superfield of the super-vielbein. 
One can note that a combination $\sim\l^2\Psi\Phi^2$ has the correct
dimension and ghost number to be an interaction term in the action
(one can also imagine a term $\sim\l^2\Phi^5$, but it will be ruled
out by gauge invariance). The main result of this paper is the
reduction of the question of 3-point couplings to the problem of
finding the operator $R^a$ relating the two fields as $\Phi^a=R^a\Psi$
and the construction of this operator. When $R^a$ represents an
operator cohomology, the interaction is non-trivial and the BV master
equation is satisfied to this 
order in the fields. 

The organisation of the paper is as follows. In Section 2, we discuss
properties of pure spinors in $D=11$. We introduce non-minimal pure
spinors, construct a regularised integration measure and discuss
convergence of integrals. There are some principal differences from
the ``standard'' case of 10-dimensional pure spinors. In section 3, we
review the known ``3-form'' and ``vielbein'' cohomologies in $\Psi$
and $\Phi^a$, respectively. We construct the operator relating the
two fields, demanding that it carries cohomology, and investigate some
other properties. Section 4 deals with
the action. We show that the master equation is satisfied to the
relevant order. In Section
5, we end with conclusions and some thoughts about the continuation of
the project, in particular higher order interactions.

\section\PureSpinors{Pure spinors in $D=11$}

\immediatesubsection\MinimalPureSpinors{Minimal pure spinors}The
anti-commutator of two fermionic derivatives in flat superspace is
typically 
$$
\{D_\a,D_\b\}=-T_{\a\b}^c\*_c=-2\g_{\a\b}^c\*_c\punkt\eqn
$$
Pure spinors are constrained by
$$
(\l\g^a\l)=0\eqn
$$ 
in order for the BRST operator
$$
q=(\l D)\eqn
$$
to be nilpotent. 

The spinors relevant for $D=11$ supergravity have 32 components. This
spinor representation is symplectic. In addition to $\e_{\a\b}$, also
$(\g^{abc})_{\a\b}$ and $(\g^{abcd})_{\a\b}$ are antisymmetric in $\a\b$, while 
$(\g^{a})_{\a\b}$, $(\g^{ab})_{\a\b}$ and $(\g^{abcde})_{\a\b}$ are
symmetric. The spinor is identical to a chiral spinor in $D=12$, where
$\e_{\a\b}$ and $\s^{(4)}_{\a\b}$ are
antisymmetric, while $\s^{(2)}_{\a\b}$ and $\s^{(5)}_{\a\b}$ are
symmetric.
Symplectic spinors, as usual, require special care with conventions to
avoid sign errors. We use conventions where indices on gamma matrices
are lowered by left and right multiplication with $\e_{\a\b}$ and
raised by its inverse. In this way the sign issues are minimised. 
We hide spinor
indices as much as possible. 
Some useful relations, both for pure and unrestricted spinors, 
are listed in Appendix A.

We call the space of $D=11$ pure spinors ${\cal P}$.
The dimension of ${\cal P}$ is 23. This can be deduced from
its decomposition in two $D=10$ spinors of opposite chirality. If we
let $\l=(\ell,m)$ the conditions become 
$$
\eqalign{
&(\ell\s^a\ell)-(m\tilde\s^a m)\komma\quad a=0\ldots9\komma\cr
&(\ell m)=0\punkt\cr
}\eqn
$$
These equations are solved by $m=iv_a\s^a\ell$, where $v_a$ is a unit vector
orthogonal to the light-like vector $(\ell\s^a\ell)$. Since there is
an equivalence under $\delta v^a=(\ell\s^a\ell)$, it represents 7
degrees of freedom (a seven-sphere) 
in addition to the 16 in $\ell$. 
Clearly, the pure spinor has to be complex, which will also be natural
when we consider (Euclidean) integration over pure spinor variables.

In $D=10$, only one irreducible module remains in the product
of any number of pure spinors. If the Dynkin label of the
$so(10)$ module of the spinor is $(00001)$, the module of the product of
$n$ pure spinors is $(0000n)$. In contrast to this, in $D=11$ the
pure spinor bilinear already contains a 2-form and a 5-form, \ie, the
modules $(01000)$ and $(00002)$. The number of irreducible modules
increases like $\Fr n2$. The irreducible modules occurring in the product of $n$
$\l$'s are
$$
\oplus_{p=0}^{[{n\over2}]}(0,p,0,0,n-2p))\punkt\Eqn\PureSpinorModules
$$
This content (or, rather, the absence of certain modules appearing in
the $n$'th symmetric product of a spinor but not in
(\PureSpinorModules)) completely determines the zero-mode cohomology
of $q=(\l D)$. 

We will discuss cohomology in the following section,
but for the sake of defining integration, we give the table of
zero-mode cohomologies (\ie, cohomology of $q$ in a field
$\Psi(\l,\th)$) already now. Calculation of the zero-mode cohomology
is a purely algebraic problem. It can be done by hand, using the
reducibility of the pure spinor constraint as in
refs. [\BerkovitsParticle,\BerkovitsICTP,\BerkovitsNekrasovCharacter], or by
computer-aided counting of modules as in ref. [\SpinorialCohomology]. 
We denote modules by
Dynkin label, and the cohomology is given in Table 1. 
We will comment on the cohomology in the following section. For now,
we will just use one of its components.

Since there are no singlets in the expansion (\PureSpinorModules)
except the constant mode, any function of $\l$ must be expressed as a
sum of positive powers. All cohomology of a scalar field comes at
$\l^m\theta^n$ with $m\leq7$, $n\leq9$. The ``top cohomology'' at
$\l^7\th^9$, constructed in the following section, is a
singlet. Picking this cohomology component from a pure spinor
superfield has all the correct properties for a measure (ghost number
-7, dimension 8, which makes the Lagrangian before integration $\int
d^{11}x$ have ghost number 0 and dimension 2) except that it is
degenerate. 
There is of course no non-degenerate ``residue'' measure if
no negative powers are allowed. This is remedied by the non-minimal
pure spinors.

There is a special subspace ${\cal P}_0$ of (complex) pure spinor space, where
$(\l\g^{ab}\l)=0$, so that one again gets only one irreducible module
at each power. Such a ``very pure'' spinor is a pure spinor in
$D=12$. The dimension of this subspace is 16, which is deduced \eg\ from the
usual counting in even dimensions using isotropic subspaces. The
real dimension of $SO(12)/SU(6)$ is $66-35=31$, which together with a
radius gives 32 real, or 16 complex. When integrating functions of a
pure spinor (and its complex conjugate) we need to check for
convergence not only at the origin $\l=0$ but also at this codimension
7 subspace. Some operators which we will find in the following
section, and
which will enter the action, 
are singular on ${\cal P}_0$.

\subsection\NonMinimalPureSpinors{Non-minimal pure spinors}Non-minimal
pure spinors were introduced by Berkovits in
ref. [\BerkovitsNonMinimal], with the purpose 
of formulating a non-degenerate measure for the pure spinors, so that
cohomology can be obtained from av action. They were further
elaborated on, and explained in term of \v Cech and Dolbeault
cohomology, by Berkovits and Nekrasov in refs. [\BerkovitsNekrasovMultiloop].

Instead of the BRST operator we wrote down in the previous section,
$q=(\l D)$, one considers
$$
Q=q+s=(\l D)+(r{\*\over\*\lb})\punkt\eqn
$$
Here, $\lb$ is another pure spinor, $(\lb\g^a\lb)=0$, and $r$ a
fermionic spinor obeying $(\lb\g^ar)=0$, so that the set of
constraints is preserved by $Q$. The last constraint means that also
$r$ has 23 independent components. When performing integrals,
one considers $\lb$ to be the complex conjugate of $\l$, $\lb_\a=(\l^\a)^\star$.
The cohomology of $Q$ is identical to that of $q$ [\BerkovitsNonMinimal], \ie,
representatives in cohomology classes can be chosen as independent of
$\lb$ and $r$. It is convenient to assign ghost number $-1$ and
dimension $\fr2$ to $\lb$, which leads to ghost number 2 and dimension
$\fr2$ for $r$.

Due to the reducibility of the modules of products of pure spinors,
there are two scalar invariants formed from $\l$ and $\lb$:
$$
\eqalign{
\xi&=(\l\lb)\komma\cr
\eta&=(\l\g^{ab}\l)(\lb\g_{ab}\lb)\punkt\cr
}\eqn
$$
The first one, $\xi$, is formally identical to the one in $D=10$. It
is positive semidefinite (from now on, we always consider $\lb$ as the
complex conjugate of $\l$) and vanishes only at the tip of the pure
spinor c\^one, $\l=0$. The second invariant, $\eta$, has no
counterpart for $D=10$ pure spinors. It is negative semidefinite 
and vanishes only on the codimension 7 subspace ${\cal
  P}_0$ of $D=12$ pure spinors. Fields and operators may contain
negative powers of both $\xi$ and $\eta$. If we want to separate the
behaviour at the origin and at ${\cal P}_0$, it is convenient to
consider the projective invariant $\tilde\eta=\xi^{-2}\eta$.

\subsection\Integration{Integration and singularities}The integration
measure for the non-minimal pure spinors is related to the tentative
residue considered in the previous subsection. Note that the existence
of the singlet at $\l^7\th^9$ implies that there is (at
least) one invariant tensor
$T_{(\a_1\ldots\a_7)[\b_1\ldots\b_9]}$. The number of antisymmetric
indices is the same as the number of constraints on a pure spinor
(this is a generic feature). We dualise the antisymmetric indices to
obtain
${\star}T_{(\a_1\ldots\a_7)}{}^{[\b_1\ldots\b_{23}]}$. 
If we follow the same procedure as in
$D=10$, we would define the integrations over the pure spinor
variables as
$$
\eqalign{
&[d\l](\l^7)^{\a_1\ldots\a_7}
={\star}\bar T^{\a_1\ldots\a_7}{}_{\b_1\ldots\b_{23}}
d\l^{\b_1}\ldots d\l^{\b_{23}}\komma\cr
&[d\lb](\lb^7)_{\a_1\ldots\a_7}
={\star}T_{\a_1\ldots\a_7}{}^{\b_1\ldots\b_{23}}
d\lb_{\b_1}\ldots d\lb_{\b_{23}}\komma\cr
&[dr]=\lb_{\a_1}\ldots\lb_{\a_7}
{\star}\bar T^{\a_1\ldots\a_7}{}_{\b_1\ldots\b_{23}}
{\*\over\*r_{\b_1}}\ldots{\*\over\*r_{\b_{23}}}\punkt\cr
}\eqn
$$
Together with full integration $[d\th]=d^{32}\th$, these integrations
have total dimension 8 and ghost number $-7$, as desired (these
numbers are insensitive to the assignment of dimension and ghost
number to $\lb$). We will use the notation $\int[dZ]$ for integration
over all coordinates, including $x$.

These equations are not fully defined as they stand. It turns out that
the singlet at $\l^7\th^9$ is not unique (although the cohomology
is). The left hand sides of the first two equations must be projected
to contain the same index structure as the right hand sides. This will
be the index structure contained in the singlet cohomology, and we let
$(\l^7)^{\a_1\ldots\a_7}$ denote this projection.

We now want to find the zero-mode cohomology at $\l^7\th^9$. When the
corresponding problem is addressed in $D=10$ one finds one singlet at
$\l^3\th^5$ and none in the ``surrounding'' positions $\l^2\th^6$ and
$\l^4\th^4$. One therefore knows that the singlet represents
cohomology. In $D=11$ we find 3 singlets at $\l^7\th^9$, and one each
in $\l^6\th^{10}$ and $\l^8\th^8$. We have to identify the cohomology
among the 3 singlets. Refining the analysis, we find that
$\l^6\th^{10}$ is formed through the scalar product of $\l^6$ and
$\th^{10}$ where both are projected to $(02002)$. We write this as
$(\l^6\modprod{(02002)}\th^{10})$. There is a module $(02002)$ already
at $\th^8$, so we can simplify further to 
$(\l^6\modprod{(02002)}\th^{8})(\th\th)$. In a similar fashion, the 3
combinations at $\l^7\th^9$ can be written
$(\l^7\modprod{(02003)}\th^9)$, $(\l^7\modprod{(03001)}\th^7)(\th\th)$
and $(\l^7\modprod{(03001)}\th^9)$, where factors $(\th\th)$ have been
written out when possible (there are two independent (03001)'s in
$\th^9$).
The singlet at $\l^8\th^8$ is $(\l^8\modprod{(04000)}\th^8)$. It is
clear that $(\l^7\modprod{(02003)}\th^9)$ is closed, since 
$(00001)\otimes(02003)$ does not contain $(04000)$. In fact, it is the
unique representative for the cohomology, but in order to show that we
need to verify that $q\cdot(\l^6\modprod{(02002)}\th^{8})(\th\th)$ has
no component in $(\l^7\modprod{(02003)}\th^9)$. We do this by explicit
calculation. Concretely,
$$
(\l^6\modprod{(02002)}\th^{8})(\th\th)=
(\l\g^{ab}\l)(\l\g^{cd}\l)(\l\g^{ijklm}\l)
(\th\g_{abp}\th)(\th\g_{cdq}\th)(\th\g_{ij}{}^p\th)(\th\g_{jkl}{}^q\th)(\th\th)
\punkt\eqn
$$
When acting with $q$ and looking only for components in
$(\l^7\modprod{(02003)}\th^9)$, we can discard everything except the
term where $q$ hits $(\th\th)$. We then also discard every expression
with 3 factors of $(\l\g^{(2)}\l)$, which gives a contribution to the
other two singlets (strictly speaking, also
$(\l^7\modprod{(02003)}\th^9)$ contains such terms along with terms
$(\l\g^{(2)}\l)^2(\l\g^{(5)}\l)$ in order to be irreducible, but the
latter ones cannot vanish).
The relevant part reads
$$
\eqalign{
&q\cdot(\l^6\modprod{(02002)}\th^{8})(\th\th)\cr
&\qquad=\ldots+2(\l\g^{ab}\l)(\l\g^{cd}\l)(\l\g^{ijklm}\l)(\l\th)
(\th\g_{jkl}{}^q\th)(\th\g_{abp}\th)(\th\g_{cdq}\th)(\th\g_{ij}{}^p\th)
\punkt\cr
}\eqn
$$
Here we use a relation from Appendix A to bring out indices to the
$\l$'s, $(\l\theta)(\theta\g^{klmq}\theta)
=-2(\l\g^{[kl}{}_r\theta)(\theta\g^{mq]r}\theta)
-(\l\g^{klmq}\theta)(\theta\theta)$, together with the fact that 
$(\g_m\l)_\a(\l\g^{ijklm}\l)$ can be thrown away when looking for the
coefficient of (02003), to get
$$
\ldots+4(\l\g^{ab}\l)(\l\g^{cd}\l)(\l\g^{ijkl[q}\l)(\l\g^{r]}\th)
(\th\g_{abp}\th)(\th\g_{cdq}\th)(\th\g_{ij}{}^p\th)(\th\g_{klr}\th)
\komma\eqn
$$
which vanishes for symmetry reasons.

So, the cohomology is uniquely represented by
$(\l^7\modprod{(02003)}\th^9)$
and takes the explicit form
$$
(\l^7\modprod{(02003)}\th^9)
=(\l\g^{ab}\l)(\l\g^{cd}\l)(\Lambda^{ijklm}\g^n\th)(\th\g_{abp}\th)(\th\g_{cdp}\th)(\th\g_{ijm}\th)(\th\g_{kln}\th)
\eqn
$$
where $\Lambda^{ijklm}_\a$ is in (00003),
$\Lambda_\a^{ijklm}=\l_\a(\l\g^{ijklm}\l)-2(\g^{[ijk}\l)_\a(\l\g^{lm]}\l)$.
This defines the tensor $T_{\a_1\ldots\a_7,\b_1\ldots\b_9}$ used in
defining the integration measure. 

The last thing to do is to regularise the integration. The measure
alone does not work properly for at least two reasons. The bosonic
integration is non-compact, and has to be regularised if the integral
of the singlet 
cohomology 
at $\l^7\th^9$ is to be finite. At the same time one needs to get a
non-zero result from the full $\th$-integration, so there must be some
factor saturating the integral with 23 $\theta$'s. All these demands
are reached by the same regularisation, which of course has to be
BRST-invariant. We
insert a factor $e^{\{Q,\chi\}}$, for some fermion $\chi$ 
[\MarneliusOgren,\BerkovitsNonMinimal]. This
differs from 1 by a $Q$-exact expression. Therefore, the integration
is independent of the choice of $\chi$, when it is
well-defined. Choosing $\chi=(\lb\theta)$ gives
${\{Q,\chi\}}=-(\l\lb)-(\theta r)$. At the same time as the bosonic
integral becomes exponentially convergent at infinity, the final term
in the expansion in $r$, $r^{23}\th^{23}$, saturates the $\th$
integration with the 23 
missing $\th$'s in the right tensorial structure to pick up the
singlet cohomology at $\l^7\th^9$.

There are possibilities of divergences both at the origin and at
${\cal P}_0$. Consider first the origin, and let
$\rho=\sqrt{(\l\lb)}$. The radial integration contains $\int
d\rho\rho^{45}$. Take an integrand $\sim\l^{p+7}\lb^p$. The integral
is convergent if $p>-23$. Close to ${\cal P}_0$, the radial coordinate
$\s$ is given by $\s^2\sim\tilde\eta$. The real codimension is 14, so
one gets an integration $\int d\s\s^{13}$. Each factor of
$(\l\g^{(2)}\l)$ or $(\lb\g^{(2)}\lb)$ goes like $\s$. The
integration measure takes away two factors of $\s$. An integrand
that behaves like $\s^q$ will give a convergent integral if 
$q>-12$.

\section\Cohomologies{Supergravity cohomologies and their relation}

\immediatesubsection\TwoCohomologies{The vielbein and 3-form
 fields}The zero-mode cohomologies in $\Psi$ and $\Phi^a$ were calculated in
ref. [\SpinorialCohomology]. 
We have listed them in Tables 1 and 2 in Appendix B. We note
that the linearised supergravity fields are obtained in both fields at
ghost number 0. The full cohomology can be understood by noting that
if there is cohomology in the next column to the right, these will
impose differential constraints on the fields. These antifield
cohomologies are in one-to-one correspondence with the equations of
motion. It is typical for maximally supersymmetric models that the
antifields are present as cohomologies of the same field as the
physical fields, so that these will make up on-shell multiplets.

In Table 1, the cohomology of $\Psi$ contains all the ghosts and
higher order ghosts relevant for the tensor gauge symmetries and
superdiffeomorphisms. 

Table 2 gives the cohomology in $\Phi^a$. It is essential that one in
addition to the pure spinor constraint consider $\Phi^a$ in the gauge
equivalence class $\Phi^a\approx\Phi^a+(\l\g^a\rho)$ for any spinor
$\rho$, otherwise the cohomology would just be the tensor product of
the cohomology in $\Psi$ with the vector module. Note that the
3-form potential $C$ only enters this cohomology through its field
strength 4-form $H=dC$. 

There are some zero-mode cohomology at ghost number zero related to
the Weyl connections [\HoweWeyl], which have no local degrees of freedom.

\vfill\eject

\subsection\RelatingCohomologies{Relating the two fields}We want to
relate the field $\Phi^a$ to the field $\Psi$ through an operator
$R^a$ of ghost number $-2$ and dimension 2. This should be possible
since they represent the same physical degrees of freedom. It will of
course not mean that any cohomology in $\Psi$ will map to a cohomology
in $\Phi^a$. The $C$-field gauge modes, the tensor ghosts and their
antifields should of course be annihilated. Neither should it be
possible to map something to any cohomology in $\Phi^a$, there are
cohomologies at negative ghost number in $\Phi^a$ which do not seem to
have any physical meaning. It is obvious that $\Psi$ should be the
fundamental field, since it allows a free action (see the following
section) and since it encodes the full set of BV fields with a
symmetry between fields and antifields. Notice for example that it is
impossible to get the Chern--Simons term from $\Phi$ alone.

The
exact form of the operator $R^a$ is not {\it a
priori} obvious. Finding good operators with negative ghost number is
non-trivial. An example of this is the $b$-ghost in $D=10$
[\BerkovitsNonMinimal], which 
contains negative powers of $\xi=(\l\lb)$.

Somewhat surprisingly, the operator $R^a$ will not contain inverse
powers of $\xi$, but of
$\eta=(\l\g^{ij}\l)(\lb\g_{ij}\lb)$. Its $r$-independent part is
$$
R_0^a=\eta^{-1}(\lb\g^{ab}\lb)\*_b\punkt\Eqn\LowestR
$$
It is clear that this operator represents non-vanishing cohomology of
$q$.
It was not initially clear that this had to be the form of $R_0^a$. If
one believes that it should capture the relation between some zero-mode
cohomology in $\Phi^a$ to
the derivative of a component field in some zero-mode position in
$\Psi$, as $H$ from $C$, it seems good. But if one focuses \eg\ on the
behaviour of the lowest component of the entire $\Phi^a$, the
diffeomorphism ghost, which comes
at $\l^2\th^2$ in $\Psi$, one might guess an expression containing two
antisymmetrised
fermionic derivatives (we will nevertheless show below how the correct
result is produced). We have shown that such an $R_0^a$ (which is
not cohomologically equivalent to the form given here) is not
possible. Eq. (\LowestR) was then obtained uniquely from the demand
that the cohomology of $f(\xi,\eta)(\lb\g^{ab}\lb)\*_b$ be independent
of $\lb$ (where $f$ obviously has to be homogeneous of degree $-2$ in $\lb$).

Closedness with respect to $Q$ means that the $q$-cohomology of $R_0^a$ 
is independent of $\lb$. We need to find a sequence of operators
$\{R_p^a\}_{p=0}^P$ of degree of homogeneity $p$ in $r$, such that
$$
\eqalign{[q,R_0^a]&=0\komma\hfill\cr
        [s,R_p]+[q,R_{p+1}^a]&=0\komma\qquad p=1,\ldots,P-1\komma\cr
        [s,R_P^a]&=0\punkt\cr
}
\eqn
$$
We get 
$$
[s,R_0^a]=r^\a{\*R_0^a\over\*\lb^\a}
=2\eta^{-1}(\lb\g^{ab}r)\*_b
-2\eta^{-2}(\lb\g^{ab}\lb)(\lb\g^{cd}r)(\l\g_{cd}\l)\*_b\punkt
\eqn
$$
This expression is $q$-exact, which can be seen by calculating
$$
\eqalign{
&[q,(\lb\g^{ab}\lb)(\lb\g^{cd}r)(\l\g_{bcd}D)]
=2(\lb\g^{ab}\lb)(\lb\g^{cd}r)(\l\g^i\g_{bcd}\l)\*_i\cr
&\qquad=2(\lb\g^{ab}\lb)(\lb\g^{cd}r)(\l\g_{cd}\l)\*_b
+4(\lb\g^{ab}\lb)(\lb\g^{cd}r)(\l\g_{bc}\l)\*_d\cr
&\qquad=2(\lb\g^{ab}\lb)(\lb\g^{cd}r)(\l\g_{cd}\l)\*_b
-2\eta(\lb\g^{ad}r)\*_d\komma\cr
}\eqn
$$
where eq. (A.{\old4}) from Appendix A has been used in the last step.
We therefore have
$$
R_1^a=\eta^{-2}(\lb\g^{ab}\lb)(\lb\g^{cd}r)(\l\g_{bcd}D)\punkt\eqn
$$
To next order we have
$$
\eqalign{
[s,R_1^a]&=
2\eta^{-2}(\lb\g^{ab}r)(\lb\g^{cd}r)(\l\g_{bcd}D)\cr
&+4\eta^{-3}(\lb\g^{ab}\lb)(\lb\g^{cd}r)(\lb\g^{ef}r)(\l\g_{ef}\l)(\l\g_{bcd}D)
\punkt\cr}\Eqn\SROne
$$
In order for this expression to be cancelled by $[q,R_2^a]$, where
$R_2^a$ is gauge invariant, it is necessary to rewrite it in a form
where the fermionic derivatives only occur through the combinations
$(\l D)$ and $(\l\g_{ij}D)$. This is possible, using identities from
Appendix A, for the expression $(\l\g_{e[f}\l)(\l\g_{bcd]}D)$. Using
this combination multiplying
$(\lb\g^{ab}\lb)(\lb\g^{cd}r)(\lb\g^{ef}r)$ as in eq. (\SROne), the
terms $(\l\g_{e[c}\l)(\l\g_{d]bf}D)$ in the antisymmetrisation will
not contribute, since they are symmetric under the interchange of the
pairs $[cd]$ and $[ef]$. With
$$
R_2^a=-16\eta^{-3}(\lb\g^{ab}\lb)(\lb\g^{cd}r)(\lb\g^{ef}r)
        (\l\g_{e[f}\l)(\l\g_{bcd]}w)\eqn
$$
we get
$$
\eqalign{
[q,R_2^a]&=
-16\eta^{-3}(\lb\g^{ab}\lb)(\lb\g^{cd}r)(\lb\g^{ef}r)
        (\l\g_{e[f}\l)(\l\g_{bcd]}D)\cr
&=-4\eta^{-3}(\lb\g^{ab}\lb)(\lb\g^{cd}r)(\lb\g^{ef}r)
\left[(\l\g_{ef}\l)(\l\g_{bcd}D)-(\l\g_{eb}\l)(\l\g_{fcd}D)\right]\cr
&=-[s,R_1^a]\punkt\cr}\eqn
$$
It is also straightforward to show that $[s,R_2^a]=0$. 
The complete operator 
$$
\eqalign{
R^a&=R_0^a+R_1^a+R_2^a\cr
&=\eta^{-1}(\lb\g^{ab}\lb)\*_b
+\eta^{-2}(\lb\g^{ab}\lb)(\lb\g^{cd}r)(\l\g_{bcd}D)\cr
&-16\eta^{-3}(\lb\g^{a[b}\lb)(\lb\g^{cd}r)(\lb\g^{e]f}r)
        (\l\g_{fb}\l)(\l\g_{cde}w)\punkt\cr}\Eqn\ROperator
$$ 
satisfies $[Q,R^a]=0$.

A further property of the operator $R^a$ is that it commutes with the
regularisation factor in the measure. It is straightforward to check
that $[R^a,(\lb\theta)]=0$. This means that $R^a$, containing only
terms with one derivative, can be partially integrated freely.

\subsection\ExampleSectionOne{An example: The diffeomorphism ghost}It
is generically technically complicated to extract components. We will
illustrate the action of the operator we have found on a specific
component in the component of the cohomology in $\Psi$, in order to
demonstrate that it really gives the correct relation between $\Phi^a$
and $\Psi$. The simplest
cohomology that is not expected to be annihilated is that of the
diffeomorphism ghost. The zero-mode sits in $\Psi$ at $\l^2\th^2$. We
will choose
$$
\Psi_\xi=(\l\g_{ij}\l)(\th\g^{ijk}\th)\xi_k(x)+\ldots\Eqn\PsiOfXi
$$ 
(one may equally well choose the structure
$(\l\g^{ijklm}\l)(\th_{ijkl}\th)\xi_m$; they differ by a $q$-exact term 
$q\cdot(\l\g^i\th)(\th\th)\xi_i$). In order to represent cohomology,
the diffeomorphism ghost must fulfill $\*_{(i}\xi_{j)}=0$, so what
remains are the ghosts corresponding to isometries (in the flat
Minkowski space we start from, translations and Lorentz
rotations). Equation (\PsiOfXi) should be complemented with a term
$\l^2\th^4\*\xi$, but it is annihilated by the derivative in
$R_0^a$. We now act with $R_0^a$ and obtain
$$
R_0^a\Psi_\xi=\eta^{-1}(\lb\g^{am}\lb)(\l\g_{ij}\l)(\th\g^{ijk}\th)\*_m\xi_i
\punkt\eqn
$$
This expression should represent a cohomology which is independent of
$\lb$, and to match the position of the same field in $\Phi^a$ it
should be proportional to
$$
\Phi^a_\xi=\xi^a+\fr4(\th\g^{aij}\th)\*_i\xi_j
$$
in cohomology. This form of $\Phi^a_\xi$ is easily checked using the
gauge symmetry $\Phi^a\approx\Phi^a+(\l\g^a\rho)$ and antisymmetry of 
$\*_i\xi_j$.
To show that $R_0^a\Psi_\xi$ captures this cohomology, we add exact
terms to $R_0^a\Psi_\xi$. A calculation shows that
$$
\eqalign{
R_0^a\Psi_\xi&+q\cdot\Bigl(\eta^{-1}(\lb\g^{am}\lb)
\bigl[16(\l\g_{mi}\th)\xi^i\cr
&\qquad-4(\l\g^i\th)(\th\th)\*_i\xi_m
+8(\l\g_m{}^{ij}\th)(\th\th)\*_i\xi_j\cr
&\qquad+4(\l\g_{mk}\th)(\th\g^{ijk}\th)\*_i\xi_j
-4(\l\g^i{}_k\th)(\th\g_m{}^{jk}\th)\*_i\xi_j
\bigr]\Bigr)\cr&=-8\Phi^a_\xi\punkt\cr
}\eqn
$$  
The first term acted on by $q$ is the only one contributing to the
zero-mode, and its coefficient is fixed by demanding that all
$\lb$-dependence disappears. The trick is to demand that after Fierz
rearrangements, only terms with $(\lb\g^{am}\lb)(\l\g_{mn}\l)$
remain, since this gives $-{1\over2}\d^a_n\eta$ using the gauge
invariance of $\Phi^a$. Although it is obvious from the construction of $R^a$
that it maps cohomology to cohomology, at least for some fields, it is
good to verify this in a concrete case.

\section\Action{The action}

\immediatesubsection\LinearisedAction{The linearised action}When there
is a non-degenerate measure, allowing partial integration by $Q$, the
equation of motion $Q\Psi=0$ is obtained from an action
$$
S_0=\int[dZ]\Psi Q\Psi\Eqn\FreeAction
$$
(here we have suppressed an overall dimensionful constant $G^{-1}$,
which is uninteresting at this stage, since all terms in the expansion
will carry the same factor).  

In a Batalin--Vilkovisky framework, the consistency criterion,
generalising $Q^2=0$ and encoding 
invariance as well as gauge algebra, is the master equation,
$$
(S,S)=0\punkt\eqn
$$
Here, the antibracket is defined as 
$$
(A,B)=\int
A{\larrowover\d\over\d\Psi(Z)}[dZ]{\rarrowover\d\over\d\Psi(Z)}B
\punkt\eqn
$$
It is a fermionic operation, and symmetric under the interchange of
bosonic $A$ and $B$. It has dimension 
$-\hbox{dim}(\int[dZ])-2\hbox{dim}(\Psi)=D-2$
and ghost number 
$-\hbox{gh\#}(\int[dZ])-2\hbox{gh\#}(\Psi)=1$.
We note that the antibracket, which in general contains a symmetrised
sum of derivatives with respect to all fields and corresponding
antifields, takes an extremely simple form.
The master equation is trivially fulfilled by the free action $S_0$.


\subsection\ThreePointCoupling{The 3-point coupling}When introducing
interactions as deformations of the the free action we let
$S=S_0+S_1+\ldots$. The master equation to lowest order reads
$(S_0,S_1)=0$. The deformation is non-trivial if 
$S_1\neq(S_0,T_1)$.
We will propose a 3-point coupling $S_1$. 

As mentioned
in the Introduction, an expression $\int[dZ]\l^2\Psi\Phi^2$ has the
correct dimension and ghost number. We will now be more specific. In
the previous section, we showed that $\Phi^a=R^a\Psi$, where the
operator $R^a$ of dimension 2 and ghost number $-2$ is given by eq.
(\ROperator). In order for $R^a$ to represent cohomology, it was
essential that $\Phi^a$ has the additional gauge invariance
$\Phi^a\approx\Phi^a+(\l\g^a\rho)$ for an arbitrary $\rho(Z)$. We
should also remember that $\Phi^a$ is a fermionic field, so any
expression like $\Phi^a\Phi_a$ vanishes; instead we need to contract
indices by some antisymmetric matrix. Both these requirements, in
addition to those from ghost number and dimension, are met
by the insertion of a factor $(\l\g_{ab}\l)$. This realisation was
inspired by the similar form of the matter kinetic term in the $D=3$
conformal theories [\CederwallBLG,\CederwallABJM]. The candidate interaction
term is
$$
S_1=\int[dZ](\l\g_{ab}\l)\Psi R^a\Psi R^b\Psi\punkt\Eqn\ThreePointInt
$$
It is invariant under gauge transformations of $\Phi^a$ 
thanks to the pure spinor
Fierz identity $(\g^b\l)_\a(\l\g_{ab}\l)=0$.
Remember that partial integration of $R^a$ is allowed. Since
$\Psi^2=0$, partial integration gives back the same expression. The
naive calculation is extremely simple. Using the expression for the
antibracket, one immediately sees that the condition for the master
equation to be fulfilled at this order is that $R^a$ is $Q$-closed,
and the condition that the interaction is non-trivial becomes the
statement that $R^a$ is not exact. We have already shown that this is
the case.

It should be mentioned that the other candidate deformation of the
action matching dimension and ghost number, 
$\int[dZ](\l\g_{abcde}\l)R^a\Psi R^b\Psi R^c\Psi R^d\Psi R^e\Psi$,
fails to make sense because it is not gauge invariant.

Before trusting the naive formal calculation, one should check that
there are no divergences neglected in the procedure. The most singular
term in $R^a$ goes as $\s^{-4}$ close to the subspace ${\cal
  P}_0$. For a non-singular $\Psi$, the integrand goes as $\s^{-7}$ or
slower, so the integral converges. This shows that $S_1$ is a valid
3-point coupling.

\subsection\ExampleTwo{Example of a coupling: The diffeomorphism ghosts}Even 
though it has been shown that the 3-point coupling
constructed is a non-trivial parameter-free deformation of the free
action, and thus must represent interacting supergravity, the
construction has been made in a rather abstract way. This is indeed
the reason that the calculations above are tractable; once physical fields
are extracted things tend to become much more
complicated. Nevertheless we would like to demonstrate that the
expected interactions arise. The example we have chosen is 
the
coupling of diffeomorphism ghosts with their antifields, which would
show that the diffeomorphism algebra is deformed from the abelian
algebra of the non-interacting theory in the way appropriate for
gravity (which is a cohomologically unique deformation [\BoulangerUniqueness]).
Remember that the gauge algebra is reflected in a coupling 
$f^a{}_{bc}c^{\star}_ac^bc^c$, where $c$ is the ghost and $c^{\star}$
its antifield.
An interesting alternative
would be to  
derive the Chern--Simons term $\int C\wedge H\wedge H$, where
$H=dC$. This is possible but quite involved, since the zero-modes of
$C$ in $\Psi$ and of $H$ in $\Phi^a$ are linear combination of a
number of terms, and the projection of products of these on the
measure mode at $\l^7\th^9$ is quite non-trivial.

The relevant terms in the fields are 
$\Psi\sim\l^2(\th^2\xi+\th^4\*\xi)+\l^5(\th^7\axi+\theta^9\*\axi)$,
$\Phi\sim\xi+\theta^2\*\xi+\l^3(\th^5\axi+\theta^7\*\axi)$. 
The coupling term in the Lagrangian
is $\l^2\Psi\Phi^2$ which then must be formed to get the singlet at
$\l^7\th^9$. We use the possibility to partially integrate $R^a$ to
make the choice to have $\axi$ in $\Psi$ and $\xi$ in $\Phi$.
Using the symmetry between the zero-mode cohomologies at
$\l^m\th^n$
and $\l^{7-m}\th^{9-n}$ we get a term with the structure
$$
\eqalign{
&\axi_i(w\g_{jk}w)(D\g^{ijk}D)\cdot(\l\g_{ab}\l)\xi^a(\th\g^{blm}\th)\*_l\xi_m\cr
&\sim\axi_i\xi^j\*_j\xi^i\cr
}\Eqn\GhostDeform
$$
with a non-zero coefficient. We have not included the
term with $\*\axi$ and two $\xi$'s, due to the technical difficulty of
writing the $\*\axi$ term in $\Psi$, but it has to contribute to the
same structure. The notation in eq. (\GhostDeform) is a little sloppy,
the $w$'s should really be the gauge-covariant derivative that
preserve the pure spinor constraint, but acting on products of $\l$'s
in modules $(0,a,0,0,b)$, \ie, in a gauge $(w\g^a w)\Psi=0$, they acts as
$w$.  

This indicates that the diffeomorphism algebra is 
obtained in the right way, although only Killing
vector ghosts were included here. 
This deformation is of course accompanied by the appropriate
interactions of the physical fields. 
We hope that this will be
convincing evidence that the proposed 3-point coupling indeed gives
couplings in $D=11$ supergravity.

\section\Conclusions{Conclusions}We have constructed a very simple
3-point coupling in an action for $D=11$ supergravity with a scalar
superfield displaying
manifest supersymmetry. All 3-point couplings between component fields
(and ghosts and antifields) are encoded in a single term. 

It will be interesting to investigate how this action continues at
higher order. A few things can be said more or less directly. When one
goes to higher couplings the coincident singularities will need to be
regularised. Hopefully this can be achieved using a similar
BRST-invariant smearing technique as in
ref. [\BerkovitsNekrasovMultiloop]. 

If
we for the moment ignore this issue, and consider the term $(S_1,S_1)$
in the master equation, it will give one term
$\sim\int(\l\g_{ab}\l)(\l\g_{cd}\l)R^a\Psi R^b\Psi R^c\Psi R^d\Psi$,
which vanishes thanks to the pure spinor constraint. There will also be a
term
$\sim\int(\l\g_{cd}\l)\Psi
R^a((\l\g_{ab}\l)R^b\Psi)R^c\Psi R^d\Psi$. It is easy to check that
even though $[R^a,(\l\g_{ab}\l)]\neq0$ one has
$[R^a,(\l\g_{ab}\l)]R^b=0$, so the remainder is
$\sim\int(\l\g_{ab}\l)(\l\g_{cd}\l)\Psi[R^a,R^b]\Psi R^c\Psi R^d\Psi$. 
The algebraic properties of $R^a$ become important. If the
commutator (which is clearly non-zero) is reasonably simple, there may
be hope of 
finding concrete forms for higher order interactions. We hope to be
able to continue along this line of pursuit. 

There is of course also a number of other questions. The formalism
suffers from a lack of background invariance. Can this is some way be
remedied? How are the conclusions altered in other backgrounds than
flat space? Can U-duality be incorporated in a dimensionally reduced
setting? Also, calculation of amplitudes might benefit from having
manifest supersymmetry. Path integral calculations of amplitudes
requires gauge fixing, which in the BV formalism also includes
elimination of antifields. Can this be achieved with a
composite ``$b$-ghost'' along the same lines as in
pure spinor string theory?

\acknowledgements The author would like to thank Bengt E.W. Nilsson,
Ulf Gran and Nathan Berkovits for discussions and comments. Special
thanks to Ulf Gran for help with GAMMA [\GAMMA].

\vfill\eject

\refout

\vfill\eject

\appendix{Spinor and pure spinor identities in $D=11$}We will list
some identities that have been useful for calculations.

Fierz rearrangements are always made
between spinors at the right and left of two spinor products. The
general Fierz identity reads
$$
(AB)(CD)=\sum\limits_{p=0}^5
\Fr{1}{32\,p!}(B\g^{a_1\ldots a_p}C)(A\g_{a_p\ldots a_1}D)\eqn
$$  
(with an overall minus sign if $A$ and one of $B$ and $C$ are
fermionic).
For bilinears in a pure spinor $\l$ this reduces to
$$
(A\l)(\l B)=-\fr{64}(\l\g^{ab}\l)(A\g_{ab}B)
+\fr{3840}(\l\g^{abcde}\l)(A\g_{abcde}B)\punkt\eqn
$$ 
From the constraint on the spinor $r$, $(\lb\g^a r)=0$, one derives 
$$
(\lb\g^{[ij}\lb)(\lb\g^{kl]}r)=0\punkt\eqn
$$
The gauge invariance $\Phi^a\approx\Phi^a+(\l\g^a\rho)$ implies that
$$
M^{ai}(\l\g_{bi}\l)=\fr2\delta^a_bM^{ij}(\l\g_{ij}\l)\komma\eqn
$$
where $a$ is the index carried by $\Phi^a$.

Various useful identities for a pure spinor $\l$ include
$$
\eqalign{ 
&(\g_j\l)_\a(\l\g^{ij}\l)=0\komma\cr
&(\g_i\l)_\a(\l\g^{abcdi}\l)=6(\g^{[ab}\l)_\a(\l\g^{cd]}\l)\komma\cr
&(\g_{ij}\l)_\a(\l\g^{abcij}\l)=-18(\g^{[a}\l)_\a(\l\g^{bc]}\l)\komma\cr
&(\g_{ijk}\l)_\a(\l\g^{abijk}\l)=-42\l_\a(\l\g^{ab}\l)\komma\cr
&(\g_{ij}\l)_\a(\l\g^{abcdij}\l)=-24(\g^{[ab}\l)_\a(\l\g^{cd]}\l)\komma\cr
&(\g_i\l)_\a(\l\g^{abcdei}\l)=\l_\a(\l\g^{abcde}\l)
-10(\g^{[abc}\l)_\a(\l\g^{de]}\l)\komma\cr}\eqn
$$
and for a fermionic spinor $\th$:
$$
\eqalign{
&\theta_\a(\theta\g^{abcd}\theta)
=-2(\g^{[ab}{}_i\theta)_\a(\theta\g^{cd]i}\theta)
-(\g^{abcd}\theta)_\a(\theta\theta)\komma\cr
&(\g_{ij}\theta)_\a(\theta\g^{aij}\theta)
=6(\g^a\theta)_\a(\theta\theta)\punkt\cr
}\eqn
$$

\vfill\eject

\appendix{Tables of cohomologies}\vskip-.4cm\noindent The horizontal
direction is the expansion in $\l$, \ie, in decreasing ghost number of
the component fields, and the vertical is the expansion of the
superfields in terms of $\th$ (downward). The columns have been
shifted in order to place fields of same dimension on the same row.

\vtop{
\vskip-.8cm
$$\hskip-2cm
\vtop{\baselineskip25pt\lineskip0pt
\ialign{
$\hfill#\quad$&$\ss\,\hfill#\hfill\,$&$\ss\,\hfill#\hfill\,$
&$\ss\,\hfill#\hfill\,$&$\ss\,\hfill#\hfill\,$&$\ss\,\hfill#\hfill$
&$\ss\,\hfill#\hfill$&$\ss\,\hfill#\hfill$
&$\ss\,\hfill#\hfill$&$\ss\,\hfill#\hfill$&\quad#\cr
\hfill\hbox{gh\#}=&\ts3&\ts2&\ts1&\ts0&\ts-1&\ts-2&\ts-3&\ts-4&\ts-5&\cr
\hbox{dim}=-3&\,\,(00000)\,\,
		&\phantom{\,\,(00000)\,\,}&\phantom{\,\,(00000)\,\,}
		&\phantom{\,\,(00000)\,\,}&\phantom{\,\,(00000)\,\,}
		&\phantom{\,\,(00000)\,\,}&\phantom{\,\,(00000)\,\,}
		&\phantom{\,\,(00000)\,\,}&\phantom{\,\,(00000)\,\,}&\cr
        -\Fr52&\bullet&\bullet&               &&&       &\cr 
           -2&\bullet&(10000)&\bullet&       &&&       &\cr
       -\Fr32&\bullet&\bullet&\bullet&\bullet&&&       &\cr
           -1&\bullet&\bullet&\raise3pt\vtop{\baselineskip6pt\ialign{
					\hfill$#$\hfill\cr
					\ss(01000)\cr
					\ss(10000)\cr}}
			&\bullet&\bullet&&\cr
       -\Fr12&\bullet&\bullet&(00001)
				&\bullet&\bullet&\bullet&&\cr
           0&\bullet&\bullet&\bullet&\raise6pt\vtop{\baselineskip6pt\ialign{
					\hfill$#$\hfill\cr
					\ss(00000)\cr
					\ss(00100)\cr
					\ss(20000)\cr}}
				&\bullet&\bullet&\bullet&&\cr
       \Fr12&\bullet&\bullet&\bullet&\raise3pt\vtop{\baselineskip6pt\ialign{
					\hfill$#$\hfill\cr
					\ss(00001)\cr
					\ss(10001)\cr}}
				&\bullet&\bullet&\bullet&\bullet&\cr
           1&\bullet&\bullet&\bullet&\bullet&\bullet&\bullet
			&\bullet&\bullet&\bullet&\cr
       \Fr32&\bullet&\bullet&\bullet&\bullet
				&\raise3pt\vtop{\baselineskip6pt\ialign{
					\hfill$#$\hfill\cr
					\ss(00001)\cr
					\ss(10001)\cr}}
				&\bullet&\bullet&\bullet&\bullet\cr
           2&\bullet&\bullet&\bullet&\bullet
				&\raise6pt\vtop{\baselineskip6pt\ialign{
					\hfill$#$\hfill\cr
					\ss(00000)\cr
					\ss(00100)\cr
					\ss(20000)\cr}}
				&\bullet&\bullet&\bullet&\bullet&\cr
       \Fr52&\bullet&\bullet&\bullet&\bullet&\bullet&(00001)&\bullet
			&\bullet&\bullet&\cr
       	   3&\bullet&\bullet&\bullet&\bullet&\bullet
				&\raise3pt\vtop{\baselineskip6pt\ialign{
					\hfill$#$\hfill\cr
					\ss(01000)\cr
					\ss(10000)\cr}}
				&\bullet&\bullet&\bullet&\cr
       \Fr72&\bullet&\bullet&\bullet&\bullet&\bullet&\bullet
				&\bullet&\bullet&\bullet&\cr
       	   4&\bullet&\bullet&\bullet&\bullet&\bullet&\bullet&(10000)
			&\bullet&\bullet&\cr
       \Fr92&\bullet&\bullet&\bullet&\bullet&\bullet&\bullet&\bullet
			&\bullet&\bullet&\cr
       	   5&\bullet&\bullet&\bullet&\bullet&\bullet&\bullet&\bullet
				&(00000)&\bullet&\cr
}}
$$
\noindent{\it Table 1. The cohomology in $\Psi$.}
}

\vfill\eject

$$\hskip-2cm
\vtop{\baselineskip25pt\lineskip0pt
\ialign{
$\hfill#\quad$&$\ss\,\hfill#\hfill\,$&$\ss\,\hfill#\hfill\,$
&$\ss\,\hfill#\hfill\,$&$\ss\,\hfill#\hfill\,$&$\ss\,\hfill#\hfill$
&$\ss\,\hfill#\hfill$&$\ss\,\hfill#\hfill$
		&\quad#\cr
           \hfill\hbox{gh\#}= &\ts1&\ts0&\ts-1&\ts-2&\ts-3&\ts-4&\ts-5&\cr
\hbox{dim}=-1&\quad(10000)\quad
		&\quad\phantom{(00000)}\quad
		&\quad\phantom{(00000)}\quad
		&\quad\phantom{(00000)}\quad
		&\quad\phantom{(00000)}\quad
		&\quad\phantom{(00000)}\quad
		&\quad\phantom{(00000)}\quad
		&\cr
        -\fr2&(00001)&\bullet&               &&&       &\cr 
           0&\bullet&(20000)&\bullet&       &&&       &\cr
       \Fr12&\bullet&\raise3pt\vtop{\baselineskip6pt\ialign{
					\hfill$#$\hfill\cr
					\ss(00001)\cr
					\ss(10001)\cr}}
			&\bullet&\bullet&&&&\cr
           1&\bullet&\raise3pt\vtop{\baselineskip6pt\ialign{
					\hfill$#$\hfill\cr
					\ss(00010)\cr
					\ss(10000)\cr}}
			&\bullet&\bullet&\bullet&&&\cr
       \Fr32&\bullet&\bullet&\raise3pt\vtop{\baselineskip6pt\ialign{
					\hfill$#$\hfill\cr
					\ss(00001)\cr
					\ss(10001)\cr}}
			&\bullet&\bullet&\bullet&&\cr
           2&\bullet&\bullet&\raise6pt\vtop{\baselineskip6pt\ialign{
					\hfill$#$\hfill\cr
					\ss(00000)(00002)\cr
					\ss(00100)(01000)\cr
					\ss(10000)(20000)\cr}}
				&\bullet&\bullet&\bullet&\bullet&\cr
       \Fr52&\bullet&\bullet&\bullet&\bullet&\bullet&\bullet&\bullet&\cr
           3&\bullet&\bullet&\bullet&\raise6pt\vtop{\baselineskip6pt\ialign{
					\hfill$#$\hfill\cr
					\ss(00000)(00002)\cr
					\ss(00100)(01000)\cr
					\ss(10000)(20000)\cr}}
				&\bullet&\bullet&\bullet\cr
       \Fr72&\bullet&\bullet&\bullet&\raise3pt\vtop{\baselineskip6pt\ialign{
					\hfill$#$\hfill\cr
					\ss(00001)\cr
					\ss(10001)\cr}}
			&\bullet&\bullet&\bullet&\cr
           4&\bullet&\bullet&\bullet&\bullet
				&\raise3pt\vtop{\baselineskip6pt\ialign{
					\hfill$#$\hfill\cr
					\ss(00010)\cr
					\ss(10000)\cr}}
			&\bullet&\bullet&\cr
       \Fr92&\bullet&\bullet&\bullet&\bullet
				&\raise3pt\vtop{\baselineskip6pt\ialign{
					\hfill$#$\hfill\cr
					\ss(00001)\cr
					\ss(10001)\cr}}
			&\bullet&\bullet&\cr
       	   5&\bullet&\bullet&\bullet&\bullet&(20000)&\bullet&\bullet&\cr
    \Fr{11}2&\bullet&\bullet&\bullet&\bullet&\bullet&(00001)&\bullet&\cr
       	   6&\bullet&\bullet&\bullet&\bullet&\bullet&(10000)&\bullet&\cr
       \Fr{13}2&\bullet&\bullet&\bullet&\bullet&\bullet&\bullet&\bullet&\cr
}}
$$

\noindent{\it Table 2. The cohomology in $\Phi^a$.}


\end